\documentclass[letterpaper,english,aps,prl,twocolumn,showpacs,superscriptaddress,footinbib]{revtex4-1}
\usepackage[T1]{fontenc}
\usepackage[latin9]{inputenc}
\usepackage{babel}
\usepackage{amsmath}
\usepackage{graphicx}
\usepackage{esint}
\usepackage[unicode=true,pdfusetitle,
 bookmarks=true,bookmarksnumbered=false,bookmarksopen=false,
 breaklinks=false,pdfborder={0 0 1},backref=false,colorlinks=false]
 {hyperref}

\makeatletter

\pdfpageheight\paperheight
\pdfpagewidth\paperwidth

 \@ifundefined{textcolor}{}
 {%
   \definecolor{BLACK}{gray}{0}
   \definecolor{WHITE}{gray}{1}
   \definecolor{RED}{rgb}{1,0,0}
   \definecolor{GREEN}{rgb}{0,1,0}
   \definecolor{BLUE}{rgb}{0,0,1}
   \definecolor{CYAN}{cmyk}{1,0,0,0}
   \definecolor{MAGENTA}{cmyk}{0,1,0,0}
   \definecolor{YELLOW}{cmyk}{0,0,1,0}
 }

\usepackage{diagbox}
\usepackage{makecell}
\usepackage{amssymb}
\makeatother

\begin{document}

\title{}

\title{Spatial Coherence and Optical Beam Shifts}

\author{W. Löffler}

\email{loeffler@physics.leidenuniv.nl}

\affiliation{Huygens Laboratory, Leiden University, P.O. Box 9504, 2300 RA Leiden,
The Netherlands}

\author{Andrea Aiello}

\affiliation{Max Planck Institute for the Science of Light, Günther-Scharowsky-Straße
1/Bldg. 24, 91058 Erlangen, Germany}

\affiliation{Institute for Optics, Information and Photonics, Universität Erlangen-Nürnberg,
Staudtstr. 7/B2, 91058 Erlangen, Germany}

\author{J. P. Woerdman}

\affiliation{Huygens Laboratory, Leiden University, P.O. Box 9504, 2300 RA Leiden,
The Netherlands}
\begin{abstract}
A beam of light, reflected at a planar interface, does not follow
perfectly the ray optics prediction. Diffractive corrections lead
to beam shifts; either the reflected beam is displaced (spatial shift)
and/or travels in a different direction (angular shift), as compared
to geometric optics. How does the degree of spatial coherence of light
influence these shifts? Theoretically, this has turned out to be a
controversial issue. Here we resolve the controversy experimentally;
we show that the degree of spatial coherence influences the angular
beam shifts, while the spatial beam shifts are unaffected.
\end{abstract}


\maketitle

A collimated optical beam is the best experimental approximation of
a ray in geometrical optics. However, due to the wave nature of light,
beams do not behave exactly as rays, and already in the case of refraction
and reflection at planar interfaces, deviations from geometric optics
occur. Goos and Hänchen \cite{goos1947} found first experimental
proof of this: an optical beam undergoes a small parallel in-plane
(longitudinal) displacement upon total reflection. Since then, multiple
variants have been found: out-of-plane shifts such as the Imbert-Fedorov
shift \cite{fedorov1955,imbert1972} and the Spin Hall Effect of Light
\cite{bliokhprl2006,hosten2008}, angular shifts \cite{merano2009},
shifts for higher-order modes \cite{merano2010}, shifts for photonic
crystals \cite{felbacq2004}, shifts for waveguides \cite{snyder1976,tsakmakidis2007},
shifts for resonators \cite{schomerus2006}, connection between beam
shifts and weak values \cite{dennis2012,dennis2012genshifts}, and
shifts for matter waves \cite{beenakker2009,haan2010,bliokh2012}. 

Surprisingly, in spite of this large body of work, the role of the
(transverse) spatial coherence of the beam has hardly been addressed.
In the original experiment by Goos and Hänchen \cite{goos1947} a
Hg lamp was used and some degree of coherence was created in a two-aperture
set-up, but this was not analyzed %
\footnote{A simple estimation based on the van Cittert-Zernike theorem would
suggest that partly coherent light was generated in \cite{goos1947},
both $\sigma_{g}$ and $\sigma_{S}$ were of similar magnitude (a
few 100 $\mathrm{\mu m}$). %
}. So, it was not clear whether this coherence was essential or not.
Almost all modern beam shift experiments have been performed with
a single-mode laser source that has near-perfect spatial coherence,
or with an extended source filtered by a single mode fiber, which
also has very good spatial coherence \cite{merano2009}. An exception
is a recent experiment \cite{schwefel2008}, which used a light-emitting
diode (without spatial filter); the authors speculate that some non-understood
aspects of their results could be due to the lack of spatial coherence
of their source. The only theoretical papers, as far as we know, that
address these issues are \cite{simon1989,wang2008,aiello2011,wang2012,aiello2012},
where \cite{simon1989,aiello2011,aiello2012} lead to diametrically
opposed results as compared to \cite{wang2008,wang2012}. We aim
in this paper to experimentally clarify the role of spatial coherence
in beam shift experiments. In short, we find that the theoretical
analysis in \cite{simon1989} and in more general form that in \cite{aiello2011,aiello2012}
does correctly describe our results.

\bigskip{}

\begin{figure}
\includegraphics[width=1\columnwidth]{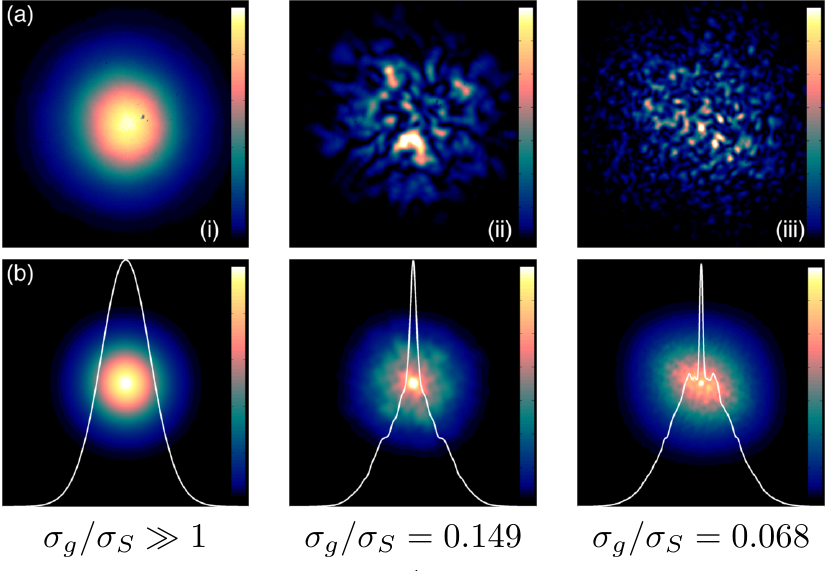} \caption{\label{fig:beams} (Color online) Gaussian Schell-model beams for
three different degrees of spatial coherence, which are used in the
experiments. The first row (a) shows exemplary intensity profiles
(in the experiment, we average over many realizations thereof by rotating
the diffuser plate). The second row (b) shows the intensity auto-correlation
of the corresponding beam in (a); as false-color plots and cross-sectional
curves. This shows clearly the two scales involved, i.e., the Gaussian
beam width $l_{S}$ and the correlation length $l_{g}$. We have determined
the ratio $l_{g}/l_{S}$ by analyzing the beam at its waist and by
measuring the far-field divergence angle \cite{mandel1995}. }
\end{figure}

We start by briefly reviewing the theory. We consider a monochromatic
partially coherent beam with a Gaussian envelope (``Gaussian Schell-model
beam'') where both the intensity distribution \emph{$I(\boldsymbol{\rho})$
and} the spatial degree of coherence $g(\boldsymbol{\rho})$ are Gaussian
\cite{martienssen1964,wolf1978,mandel1995} ($\boldsymbol{\rho}$
is the transverse position): 
\begin{equation}
I(\boldsymbol{\rho})\propto\exp\left(-\frac{\rho^{2}}{2\sigma_{S}^{2}}\right),\quad g(\boldsymbol{\rho})=\exp\left(-\frac{\rho^{2}}{2\sigma_{g}^{2}}\right)
\end{equation}
 In the source plane, $\sigma_{S}$ is the coherent (Gaussian) mode
waist, and $\sigma_{g}$ determines the correlation length. The latter
approaches infinity for a fully coherent mode, and is a measure of
the speckle size in case of partial spatial coherence. After propagating
over a distance $z$, these quantities evolve into $\sigma_{g}(z)$
and $\sigma_{S}(z)$; however, it turns out that their ratio $\sigma_{g}/\sigma_{S}$
is independent of propagation \cite{collett1980}: $\sigma_{g}(z)/\sigma_{S}(z)=\sigma_{g}/\sigma_{S}$;
therefore we use this ratio to quantify transverse coherence. Fig.~\ref{fig:beams}(a)
shows three examples of such beams, from fully coherent (i) to the
case where the coherence length is below one tenth of the beam size
(iii). By calculating the intensity autocorrelation, the two length
scales which are involved become visible: The Gaussian envelope leads
to a wide background, while the emerging speckles in case (ii) and
(iii) add a short-range correlation as is easily visible in the cross-section
curves in Fig.~\ref{fig:beams}(b). The number of participating modes
is approximately $\left(\sigma_{g}/\sigma_{S}\right)^{-2}$, which
is $(1,\approx50,\approx200)$ for the cases (i,ii,ii) respectively.
The three beams shown in Fig.~\ref{fig:beams} have been used in
the experiments reported below. 

\begin{figure}
\includegraphics[height=2.5cm]{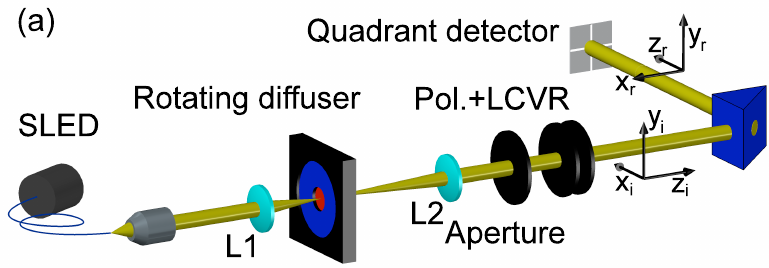}

\includegraphics[width=1\columnwidth]{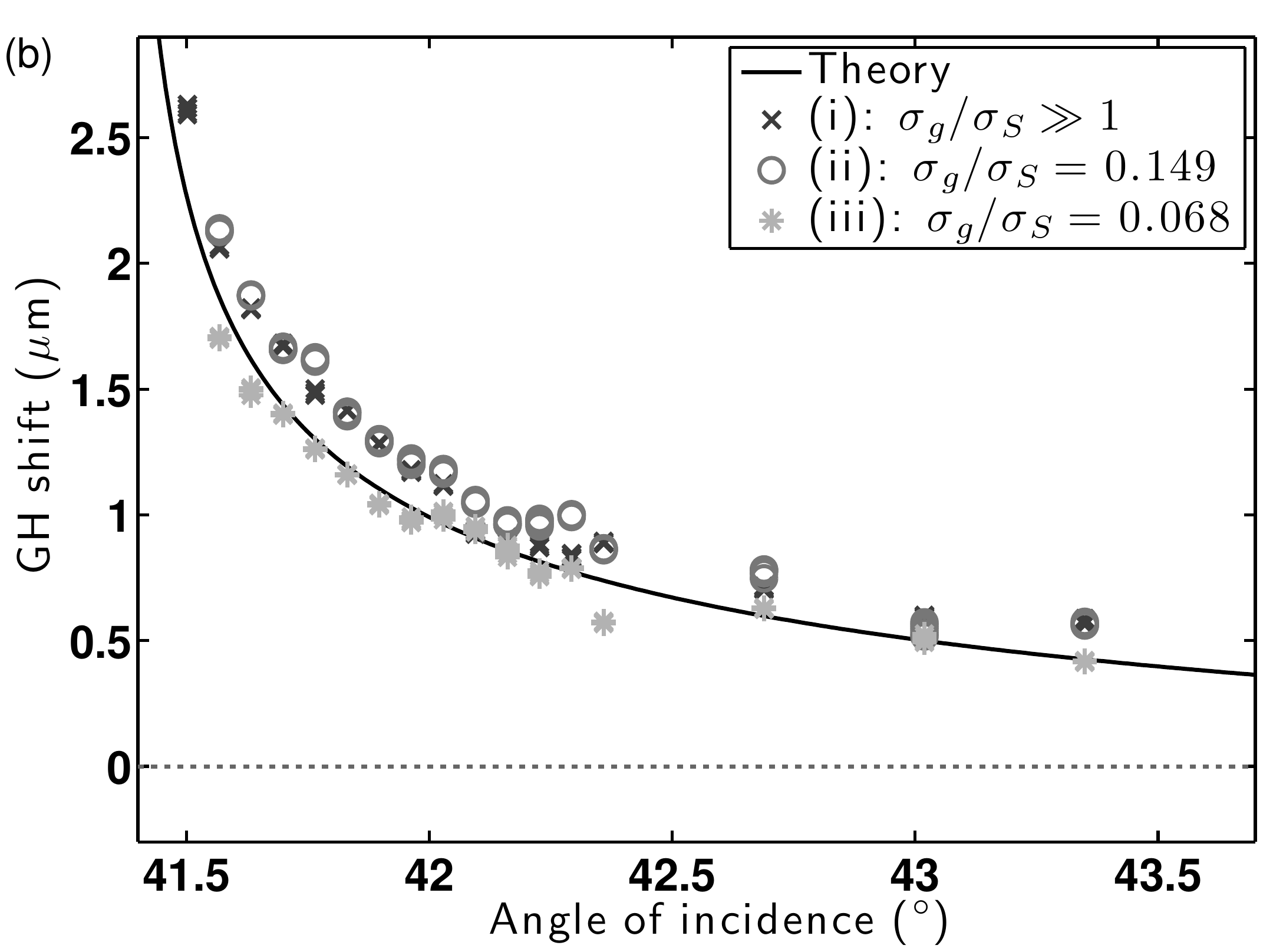}\caption{\label{fig:spgh}(Color online) Setup (a): An optical beam with variable
spatial coherence is prepared by collimating the light scattered from
a holographic diffuser plate. A combination of a polarizer and a liquid-crystal
variable retarder (LCVR) is used to modulate between $s$ and $p$
polarization. After total internal reflection from the prism, the
displacement is determined with a quadrant detector. Spatial Goos-Hänchen
shift measurements (b): Experimentally obtained spatial GH beam shifts
for different degrees of spatial coherence. Shown is the observed
polarization-differential shift (symbols), the black curve corresponds
to the theoretical prediction.}
\end{figure}

To be able to discuss such a Gaussian Schell-model beam within the
unifying beam shift framework developed by Aiello and Woerdman \cite{aiellobp2008},
we consider a paraxial, monochromatic and homogeneously polarized
($\lambda=1,2\equiv p,s$), but otherwise arbitrary, incoming optical
field $\mathbf{U}^{i}(x,y)=\sum_{\lambda}U(x,y)a_{\lambda}\hat{\mathbf{x}}_{\lambda}^{i}$.
It propagates along $\hat{\mathbf{x}}_{3}^{i}$ ($z$ coordinate),
and $\left(a_{1},a_{2}\right)$ is its polarization Jones vector.
We use dimensionless quantities in units of $1/k$, where $k$ is
the wavevector. The coordinate systems and their unit vectors $\hat{\mathbf{x}}_{\lambda}^{i,r}$
are attached to the incoming ($i$) and reflected ($r$) beam, respectively
{[}see Fig.~\ref{fig:spgh}(a){]}. After reflection at a dielectric
interface, the polarization and spatial degree of freedom are coupled
by the Fresnel coefficients $r_{p,s}$ as \cite{merano2010}

\begin{equation}
\mathbf{U}^{r}(x,y,z)=\sum_{\lambda}a_{\lambda}r_{\lambda}U(-x+X_{\lambda},y-Y_{\lambda},z)\hat{\mathbf{x}}_{\lambda}^{r}.
\end{equation}

$X_{1,2}$ and $Y_{1,2}$ are the polarization-dependent dimensionless
beam shifts:

\begin{subequations}
\begin{align}
X_{1} & =-i\,\partial_{\theta}\left[\ln r_{1}(\theta)\right],\; Y_{1}=i\frac{a_{2}}{a_{1}}\left(1+\frac{r_{2}}{r_{1}}\right)\cot\theta
\end{align}

\vspace{-5mm}

\begin{align}
X_{2} & =-i\,\partial_{\theta}\left[\ln r_{2}(\theta)\right],\; Y_{2}=-i\frac{a_{1}}{a_{2}}\left(1+\frac{r_{1}}{r_{2}}\right)\cot\theta
\end{align}

\end{subequations}

Their real parts correspond to spatial beam shifts, and their imaginary
parts to angular beam shifts. For either variant, beam displacements
$X_{\lambda}$ (along $\hat{x}$ coordinate) correspond to longitudinal
Goos-Hänchen (GH) type shifts \cite{goos1947}, while transverse displacements
$Y_{\lambda}$ along $\hat{y}$ have Imbert-Fedorov (IF) character
\cite{imbert1972,bliokhprl2006}. We observe that the transverse shifts
$Y_{\lambda}$ require simultaneously finite $a_{1}$ and $a_{2}$,
such as present in circularly polarized light; this is not necessary
for the longitudinal shifts $X_{\lambda}$. This explains why the
spatial shifts depend in the GH case only on one reflection phase,
$\phi_{1}$ or $\phi_{2}$ with $\phi_{\lambda}=\arg(r_{\lambda})$,
while in the IF case, the spatial shift depends on the phase difference
(e.g., $\phi_{1}-\phi_{2}$).

In the lab, beam shifts are usually measured via the centroid of the
reflected beam

\begin{equation}
\langle\mathbf{R}\rangle(z)=\sum_{\lambda}w_{\lambda}\frac{\int\rho\langle\left|U(-x+X_{\lambda},y-Y_{\lambda},z)\right|^{2}\rangle\mathrm{d}x\mathrm{d}y}{\int\langle\left|U(-x+X_{\lambda},y-Y_{\lambda},z)\right|^{2}\rangle\mathrm{d}x\mathrm{d}y},\label{eq:centroidshift}
\end{equation}

where $w_{\lambda}=\left|r_{\lambda}a_{\lambda}\right|^{2}\big/\sum_{\nu}\left|r_{\nu}a_{\nu}\right|^{2}$
is the fraction of the reflected intensity with polarization $\lambda$.
Eq.~\ref{eq:centroidshift} can be calculated by Taylor expansion
around zero shift ($X_{\lambda}=Y_{\lambda}=0$).  With the spatial
$\mathbf{\Delta}_{\lambda}=\mathrm{Re}\left(X_{\lambda},Y_{\lambda}\right)$
and angular $\mathbf{\Theta}_{\lambda}=\mathrm{Im}\left(X_{\lambda},Y_{\lambda}\right)$
shift vectors we obtain for the centroid $\langle\mathbf{R}\rangle(z)=\sum_{\lambda}w_{\lambda}\left(\mathbf{\Delta}_{\lambda}+M(z)\mathbf{\Theta}_{\lambda}\right)$,
where $M(z)$ is a polarization-independent $2\times2$ matrix which
couples longitudinal and transverse beam shifts depending on the transverse
mode of the field \cite{merano2010}.

For a spatially incoherent beam, the incoming field $\mathbf{U}^{i}$
corresponds to one realization of the ensemble of random fields with
equal statistical properties. For our case of a Gaussian Schell-model
beam, $M(z)$ turns out to be diagonal \cite{aiello2011}, and we
finally obtain

\begin{equation}
\langle\mathbf{R}\rangle(z)=\sum_{\lambda=1}^{2}w_{\lambda}\left[\mathbf{\Delta}_{\lambda}+\mathbf{\Theta}_{\lambda}\theta_{S}^{2}z\right].\label{eq:shiftgsm}
\end{equation}

The first term is independent of $z$, it therefore describes shifts
of purely spatial nature. Since spatial coherence enters the discussion
only via the parameter $\theta_{S}$ (which we discuss below), and
the first term is independent thereof, we conclude that spatial shifts
are expected to be independent of the degree of transverse coherence.

We test this in our first experiment {[}Fig.~\ref{fig:spgh}(a){]},
where we examine the \emph{spatial} Goos-Hänchen shift (extension
to the spatial Imbert-Fedorov case is straightforward). We collimate
light from a single-mode fiber-coupled 675~nm superluminescent diode
(FWHM spectral width 20~nm) with a $20\times$ objective and focus
it loosely ($f_{L1}=20$~cm) close to the outer edge of a holographic
diffuser (21~mm diameter, scattering angle 0.5 degree) \cite{martienssen1964}.
This diffuser is rotated at 70~Hz, which leads to a modulation in
the speckle pattern at $\sim$30~kHz (this is related to the microscopic
structure of the diffuser). This frequency is much higher than the
polarization modulation frequency (see below). We collimate the far
field ($f_{L2}=10$~cm) from the plate and use an adjustable diaphragm
{[}see Fig.~\ref{fig:spgh}(a){]} to gain full control over the key
parameter $\sigma_{g}/\sigma_{S}$. We implement polarization modulation
(10~Hz) using a polarizer in combination with a liquid-crystal variable
retarder to generate an $s$ or $p$ polarized beam. This beam is
reflected under total internal reflection in a $45^{\circ}-90^{\circ}-45^{\circ}$
prism (BK7, $n=1.514$ at 675~nm), and refraction at the side faces
of the prism is taken into account for determination of the angle
of incidence $\theta$. A quadrant detector in combination with a
lock-in amplifier (locked to the polarization modulation) is used
to measure the relative beam displacement (the quadrant detector is
binned so that it effectively acts as a binary split detector). Fig.~\ref{fig:spgh}(b)
shows the measured spatial GH shifts for the three beams with different
spatial coherence shown in Fig.~\ref{fig:beams}. We present exclusively
polarization-differential shifts $D_{ps}=D_{p}-D_{s}$, where $D_{p,s}$
are the displacements of $p$ and $s$ polarized reflected beams from
the geometrical-optics position. For $\sigma_{g}/\sigma_{S}\gg1$
we recover the well-known result that the spatial GH shift appears
only for $\theta>\theta_{c}$ \cite{artmann1948} where $\theta_{c}$
is the critical angle of $41.35^{\circ}$. However, the essential
point of Fig.~\ref{fig:spgh}(b) is, that we find that the spatial
beam shift is in fact \emph{independent} from the degree of spatial
coherence. This demonstrates that the theoretical result in \cite{simon1989,aiello2011,aiello2012}
is correct, contrary to competing claims \cite{wang2008,wang2012}.

\begin{figure}
\includegraphics[height=2.5cm]{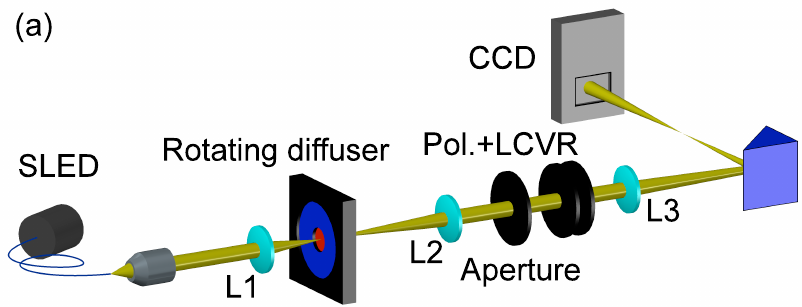}

\includegraphics[width=1\columnwidth]{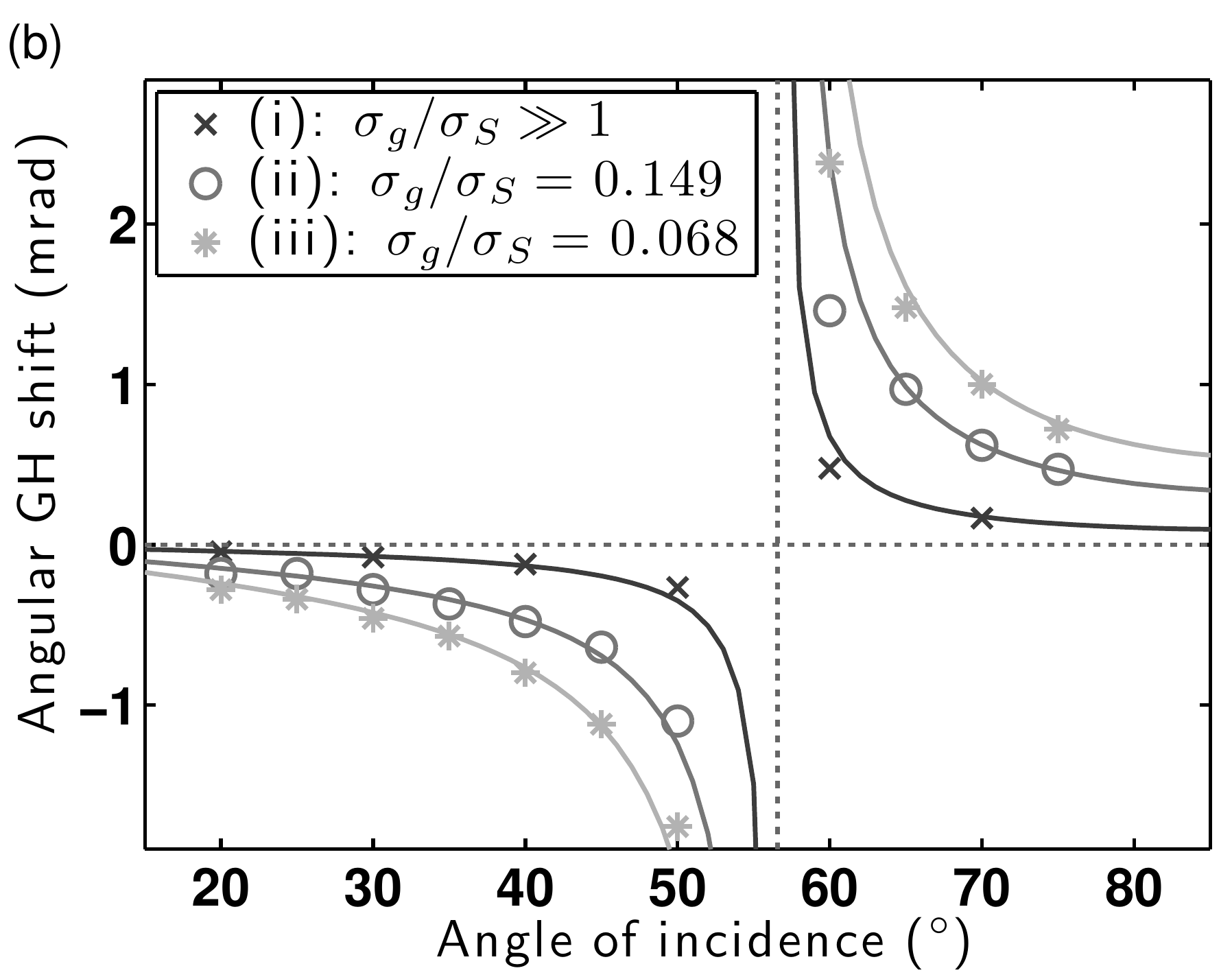}\caption{\label{fig:anggh}(Color online) Setup to measure angular shifts (a):
Compared to the experiment in Fig.~\ref{fig:spgh}(a), we introduce
lens L3 to give the beam a sizable angular spread, and we use external
reflection from the prism hypotenuse face. Further, we use a CCD camera
and centroid determination by a computer to measure the relative beam
position for $s$ and $p$ polarization. (b): Angular beam shifts
for different degrees of coherence. The experimental data (symbols)
agree with theory (curves), the vertical line indicates the Brewster
angle ($56.55^{\circ}$).}
\end{figure}

We turn now to the \emph{angular} shifts, i.e., to the second term
in Eq.~\ref{eq:shiftgsm}. The parameter $\theta_{S}$ is simply
the effective beam divergence half-angle for a Gaussian Schell-model
beam \cite{mandel1995}:

\begin{equation}
\theta_{S}^{2}=\frac{2}{k^{2}}\left[\left(\frac{1}{2\sigma_{S}}\right)^{2}+\left(\frac{1}{\sigma_{g}}\right)^{2}\right].\label{eq:theta0}
\end{equation}

We see that reduced spatial coherence, i.e., reduced $\sigma_{g}$,
leads to increased beam divergence, and this in turn leads to increased
angular beam shifts.

We test this in our second experiment, where we investigate the case
of the in-plane (Goos-Hänchen type) \emph{angular} beam shift. For
this we use, as shown in Fig.~\ref{fig:anggh}(a), an additional
lens L3 ($f_{L3}=10$~cm) to focus the beam, which is now reflected
externally at the hypotenuse plane of the same prism as used before.
The angular shift implies that $s$ and $p$ polarized beams follow
slightly different paths which both originate at the beam waist \cite{merano2009}.
For our experimental parameters and for beam propagation of a few
centimeters, this angular shift is expected to lead to many tens of
$\lambda$ displacements of the centroid. We can then use simply a
CCD camera to determine the difference in centroid position for $p$
and $s$ polarization. From two of such measurements at different
propagation distance (5~cm apart) we determine the angular Goos-Hänchen
shift, see Fig.~\ref{fig:anggh}(b). The angular shift shows a dispersive
shape around the Brewster angle $\theta_{B}$. This in itself is well
known \cite{merano2009}; it is a consequence of the fact that the
amplitude reflectivity flips sign at its zero crossing at $\theta_{B}$.
New is that we find a strong influence of the degree of coherence
on the shift, in perfect agreement with the theoretical curves shown.
We conclude that also in this case the theoretical predictions in
Refs.~\cite{simon1989,aiello2011,aiello2012} are correct. 

\begin{figure}
\includegraphics[width=1\columnwidth]{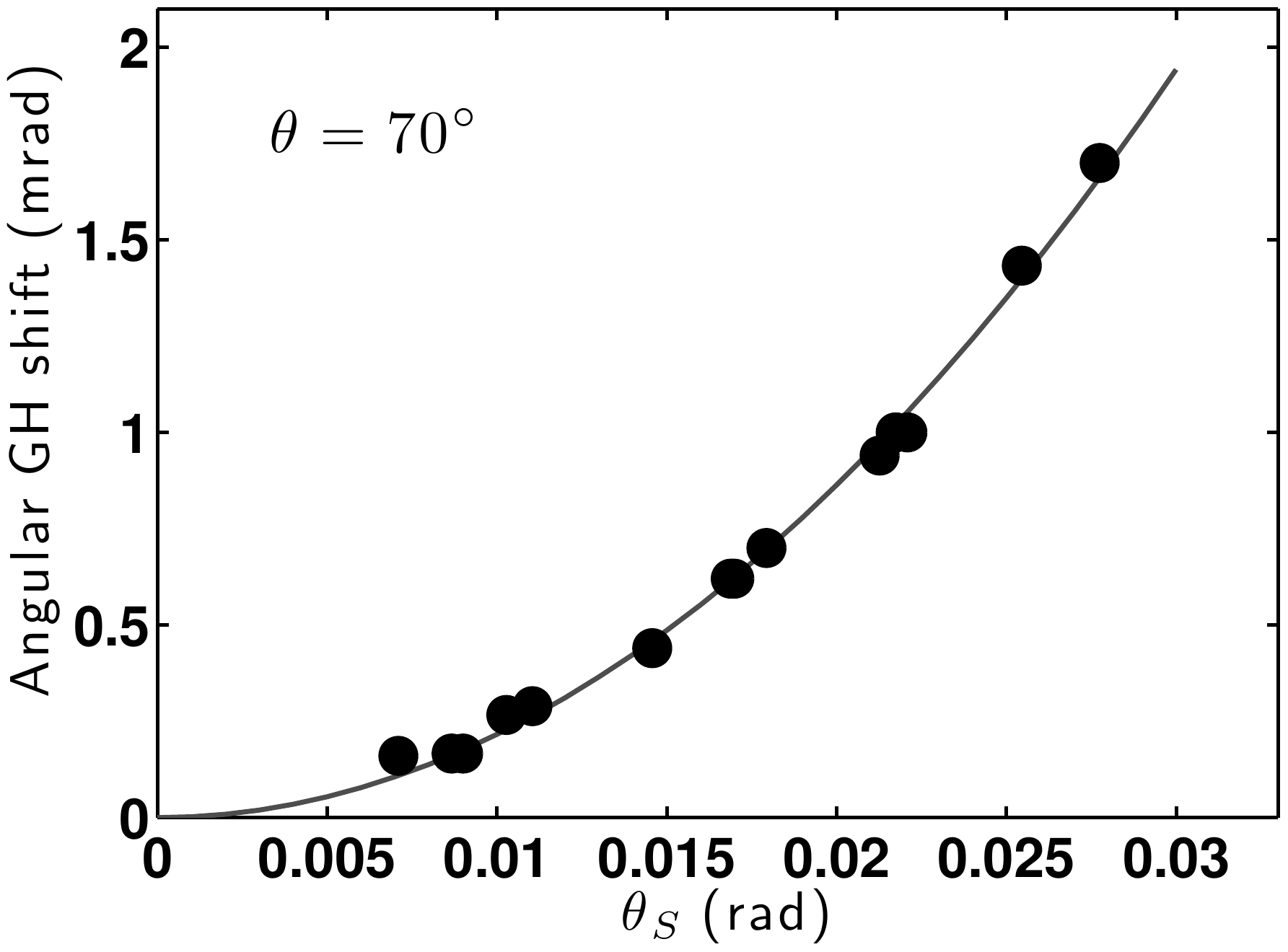}\caption{\label{fig:angghvst0}Demonstration of the particular nature of the
decoherence-enhanced angular beam shifts: The shifts scale with $\theta_{S}^{2}$.
Dots are experimental data. The measurement uncertainty is of the
order of the size of the dots. The curve shows the theoretical prediction
(there is no fit parameter involved). }
\end{figure}
Further, we note that if we replace the coherent-mode opening angle
$\theta_{0}$ in the angular GH shift formulas by the effective beam
opening angle $\theta_{S}$ (see Eq.~\ref{eq:theta0}), partially
coherent beams are well described. We therefore expect that, for a
constant angle of incidence $\theta$, the angular beam shift is proportional
to $\theta_{S}^{2}$. This we have demonstrated experimentally for
$\theta=70^{\circ}$, see Fig.~\ref{fig:angghvst0}.

In conclusion, we have found experimentally that partial spatial coherence
of a beam does not affect spatial beam shifts, while angular beam
shifts are enhanced. Basically, reduced spatial coherence increases
the effective angular spread of the beam, and therefore, angular shifts
are increased. Our data is in good agreement with the theoretical
study of Simon and Tamir \cite{simon1989}, as well as later work
\cite{aiello2011,aiello2012}. We can conclude that the dispute in
literature \cite{aiello2011,wang2012} is now definitively resolved.

We note that partially coherent beams have several advantages: they
are less vulnerable to speckle formation and also less susceptible
to atmospheric turbulences \cite{wu1991}. Although our results have
been obtained for a single dielectric interface, this can be easily
extended to the case of multilayer dielectric mirrors and metal mirrors.
Also, despite that our experimental results are for longitudinal Goos-Hänchen
type shifts only, it is clear from theory that the spatial and angular
transverse Imbert-Fedorov shifts depend in the same way on the degree
of spatial coherence as the spatial and angular GH shifts do. Our
findings demonstrate that transverse-incoherent sources, such as light-emitting
diodes, can be used in applications which use beam shifts as a sensitive
meter, such as in bio-sensing \cite{yin2006}. Finally, our findings
are relevant for beam shifts of particle beams (such as electron beams
\cite{bliokh2012} or other matter beams \cite{haan2010}). Such beams
are extremely difficult to prepare in a single mode (contrary to light
beams) due to the smallness of the De Broglie wavelength; however,
we know now that this should not diminish their beam shifts.

\begin{acknowledgments}
We acknowledge fruitful discussions with M. P. van Exter and financial
support by NWO and the EU STREP program 255914 (PHORBITECH).
\end{acknowledgments}

\begin{center}

\par\end{center}

\end{document}